\RequirePackage[l2tabu, orthodox]{nag}
\RequirePackage{snapshot}

\documentclass[9pt,twocolumn]{extarticle}
\makeatletter
\if@twocolumn
  \usepackage[dvips,letterpaper,top=0.5in, bottom=0.5in, left=0.75in, right=0.5in,includefoot,heightrounded]{geometry}
\else
  \usepackage[dvips,letterpaper,margin=1in,includefoot,heightrounded]{geometry}
\fi

\usepackage{srcltx}

\usepackage{amsmath}
\usepackage{amssymb,amsfonts}

\usepackage{abstract}

\usepackage{graphicx}
\usepackage[usenames,dvipsnames]{color}
\usepackage[usenames,dvipsnames]{xcolor}

\usepackage{booktabs}

\usepackage{setspace}
\usepackage{flushend}
\usepackage{multicol}

\usepackage{url}
\usepackage{siunitx}

\usepackage[tiny]{titlesec}
\titleformat{\section}[hang]{\filright\scshape}{}{0em}{\thesection\quad}
\titleformat{\subsection}[hang]{\filright\scshape}{}{0em}{\thesubsection\quad}
\titleformat{\subsubsection}[hang]{\filright\scshape}{}{0em}{\thesubsubsection\quad}

\titleformat{\section}[hang]{\filright\scshape}{}{0em}{\thesection\quad}

\titleformat{name=\section,numberless}[hang]
{\filright\scshape}{}{0em}{}

\newcommand{\getA}[1]{{#1}^{(a)}}
\newcommand{\getB}[1]{{#1}^{(b)}}
\newcommand{\getC}[1]{{#1}^{(c)}}
\newcommand{\getD}[1]{{#1}^{(d)}}

\makeatletter

\newcommand{\printtitle}{%
\makeatletter
\if@twocolumn

\twocolumn[%
  \maketitle
  \begin{onecolabstract}
    \myabstract
  \end{onecolabstract}
  \begin{center}
    \small
    \textbf{Keywords}
    \\\medskip
    \mykeywords
  \end{center}
  \bigskip
]
\saythanks
\else
  \maketitle
  \begin{onecolabstract}
    \myabstract
  \end{onecolabstract}
  \begin{center}
    \small
    \textbf{Keywords}
    \\\medskip
    \mykeywords
  \end{center}
  \bigskip
  \onehalfspacing
\fi
\makeatother
}

\title{%
A Single-Channel Architecture for Algebraic Integer Based
8$\times$8 \mbox{2-D} DCT Computation}

\author{%
A.~Edirisuriya%
\thanks{%
A.~Edirisuriya, A.~Madanayake and N.~T. Rajapaksha are with
Departament of Electrical and Computing Engineering,
University of Akron, Akron, OH, USA
e-mail: arjuna@uakron.edu}%
\quad
A.~Madanayake$^\ast$%
\quad
R.~J.~Cintra%
\thanks{R. J. Cintra is with the
Signal Processing Group,
Departamento de Estat\'istica,
Universidade Federal de Pernambuco,
PE 50740-540, Brazil
e-mail: rjdsc@de.ufpe.br}%
\quad
V.~S.~Dimitrov%
\thanks{V. S. Dimitrov is with the
Department of Electrical and Computer Engineering,
University of Calgary, Calgary, AB T2M 4S7, Canada
e-mail: vdvsd103@gmail.com}
\quad
N.~T. Rajapaksha$^\ast$%
}

\date{}

\newcommand{\myabstract}{%
An area efficient row-parallel architecture is proposed for the real-time implementation of bivariate algebraic integer (AI) encoded \mbox{2-D} discrete cosine transform (DCT) for image and video processing. The proposed architecture computes 8$\times$8 \mbox{2-D} DCT transform based on the Arai DCT algorithm.
An improved fast algorithm for AI based 1-D DCT computation is proposed along with a single channel 2-D DCT architecture.
The design improves on the 4-channel AI DCT architecture that was published recently
by reducing the number of integer channels to one
and the number of 8-point~\mbox{1-D} DCT cores from 5 down to 2.
The architecture offers exact computation of 8$\times$8 blocks of the
\mbox{2-D} DCT coefficients up to the FRS,
which converts the coefficients from the AI representation to fixed-point format using the method of expansion factors.
Prototype circuits corresponding to FRS blocks based on two expansion factors are realized, tested, and
verified on FPGA-chip, using a Xilinx Virtex-6 XC6VLX240T device.
Post place-and-route results show a 20\% reduction
in terms of area compared to the
\mbox{2-D} DCT architecture requiring five 1-D AI cores.
The area-time and area-time${}^2$ complexity metrics are also reduced by 23\% and 22\% respectively for designs with 8-bit input word length.
The digital realizations are simulated up to place and route
for ASICs using 45~nm CMOS standard cells.
The maximum estimated clock rate is 951~MHz
for the CMOS realizations indicating
7.608$\cdot$10$^9$~pixels/seconds and a 8$\times$8
block rate of 118.875~MHz.
}

\newcommand{\mykeywords}{%
DCT, Algebraic Integers, Expansion factors
}

\begin{document}

\printtitle

\section{Introduction}

The discrete cosine transform (DCT)
is a widely used mathematical tool in image and video compression.
It is a core component in contemporary media standards
such as JPEG and MPEG~\cite{mpeg}.
Indeed,
the DCT is known for its  properties of
decorrelation,
energy compaction, separability, symmetry,
and orthogonality~\cite{blinn}.
However the computational complexity of the DCT operation imparts
a significant burden in VLSI circuits for real time applications.
Several algorithms have been proposed to
reduce the complexity of DCT circuits
by exploiting its mathematical properties~\cite{arai,DimitrovJullienMiller1998,wahidbook}.
Integer transforms are employed in video standards such as H.264.
However, we emphasize that these methods are approximations, which are inherently inexact and may introduce a computational
error floor to the DCT evaluation.

To wit, one of the main obstacles in performing accurate DCT computations
is the implementation of the
irrational coefficient multiplications
needed to calculate the transform.
Traditional DCT implementations adopt a compromise solution
to this problem
employing
truncation or rounding off~\cite{roundoff1}
to approximate these quantities.
As a consequence,
computational errors are systematically introduced into the computation, leading to degradation of the signal-to-noise ratio~(SNR).

To partially address this issue, algebraic integer~(AI) encoding~\cite{Cozzens_1985}
has been employed~\cite{Games_1989,wahid20042D,Mohammadi_2010,Pardani_2011}.
The main idea in this approach is to map required irrational numbers
into an array of integers,
which can be arithmetically manipulated in an error free manner.
At the end of the computation,
AI based algorithms require a
final reconstruction step (FRS) in order
to map the resulting encoded integer arrays
back into the usual fixed-point representation.
FRS can be implemented by means of individualized circuits,
at in principle \emph{any given precision}~\cite{DimitrovJullienMiller1998}.

Architectures based on the low-complexity Arai DCT algorithm~\cite{arai}
has been proposed in~\cite{wahid20042D,wahid2005error}.
The Arai DCT algorithm is an algorithm for 8-point DCT computation
in video and image processing applications because of its relatively low computational complexity.
It is noted that this algorithm requires only
five multiplications to generate
the eight output coefficients.
In the AI based architectures proposed in~\cite{wahid20042D,wahid2005error},
the algebraically encoded numbers are
reconstructed and represented in fixed-point format
at the end of column-wise DCT calculation
by means of an intermediate reconstruction step;
then data is coded again
before the row-wise DCT calculation.
As a result,
the intermediate reconstruction step
introduces errors that propagate into subsequent sections.
In a sense,
the presence of such computational error
in a intermediate stage of the calculation
diminishes the point of employing an AI based structure.

To address this issue,
in~\cite{TCSVT}
a doubly AI encoded architecture
where the reconstruction is performed only once
at the end of the entire computation was proposed.
Such architecture
could allow a completely error free
computation throughout all algorithm stages
until the reconstruction stage for the \mbox{2-D} DCT computation.
In fact,
after the column-wise computation,
data path is divided and the row-wise computation
is performed separately for each AI base. We use the term channel to refer to these data paths.

In this paper,
we propose an improved fast algorithm for
the AI based \mbox{1-D} DCT computation
derived from the algorithm proposed in~\cite{wahid20042D}.
Additionally,
we present a \mbox{2-D} DCT architecture based on
the improved \mbox{1-D} transform.
This \mbox{2-D} constitutes of a single channel
which provides improvements in terms of area and power consumption
when compared with the four channel architecture
described in~\cite{TCSVT}.
We show that these improvements could be obtained
without making any compromise in terms of accuracy.
Detailed comparisons between the proposed architecture with
some of the existing are also provided.

\section{Review of AI based DCT computation}

AI encoding provides
an exact representation devoid of
quantization noise in DCT computations.
An AI is defined as a root of a monic polynomial
whose coefficients are integers~\cite{hardy1975numbers}.
AI may constitute a basis and real numbers may be expressed
as a integer linear combination of such basis elements.
Thus,
real numbers can be possibly represented
without errors
by an array of integers.

In~\cite{wahid20042D} AI encoding is adapted into Arai DCT algorithm~\cite{arai}, and the corresponding encoding is shown in Table~\ref{table.AI_Encoding}.
The corresponding AI basis is furnished by
$\left[ \begin{smallmatrix} 1 & z_1 \\ z_2 & z_1z_2 \end{smallmatrix}\right]$,
where $z_1 = \sqrt{2+\sqrt{2}} + \sqrt{2-\sqrt{2}}$
and
$z_2 = \sqrt{2+\sqrt{2}} - \sqrt{2-\sqrt{2}}$.
It should be noted that the hardware implementation of this representation requires only of adders/subtracters.

\begin{table}
\centering
\caption{\mbox{2-D} AI encoding of Arai DCT constants}
\label{table.AI_Encoding}
\begin{tabular}{c|c}
\hline
$\cos(4\pi /16)$ & $\cos(2\pi /16)-\cos(6\pi /16)$\tabularnewline
\hline
\\ [-2ex]
$\begin{bmatrix}0 & 0 \\ 0 & 1 \end{bmatrix}$ & $\begin{bmatrix}0 & 0 \\ 2 & 0 \end{bmatrix}$\tabularnewline
\\ [-2ex]
\hline

$\cos(6\pi /16)$ & $\cos(2\pi /16)+\cos(6\pi /16)$\tabularnewline
\hline
\\ [-2ex]
$\begin{bmatrix}0 & 1 \\ -1 & 0 \end{bmatrix}$ & $\begin{bmatrix}0 & 2 \\ 0 & 0 \end{bmatrix}$\tabularnewline
\hline
\end{tabular}
\end{table}

Indeed, a given real number $x$ is encoded
into an integer array
$\left[\begin{smallmatrix} \getA{x} & \getB{x} \\ \getC{x} & \getD{x}\end{smallmatrix}\right]$,
where $\getA{x}$, $\getB{x}$, $\getC{x}$, and $\getD{x}$
indicate the integers associated to basis elements
1, $z_1$, $z_2$, and $z_1z_2$, respectively.
We refer to these integers as the AI components
of $x$.

Therefore,
quantity $x$ can be decoded (reconstructed)
from its AI encoded form according to~\cite{wahid20042D}:
\begin{equation}
\label{z4.representation}
\begin{split}
x
&
\equiv
\operatorname{tr}
\left(
\begin{bmatrix}
\getA{x} & \getB{x} \\
\getC{x} & \getD{x}
\end{bmatrix}
\cdot
\left[ \begin{smallmatrix} 1 & z_1 \\ z_2 & z_1z_2 \end{smallmatrix}\right]^\top
\right)
\\
&=
\getA{x}
+
\getB{x} \cdot z_1
+
\getC{x} \cdot z_2
+
\getD{x} \cdot z_1 z_2
,
\end{split}
\end{equation}
where $\operatorname{tr}(\cdot)$ returns the trace of its argument
and superscript~${}^\top$ corresponds to the transposition operation.
The decoding can be done in a
tailored FRS where the AI basis is represented at the desired precision.
Several FRS structures have been proposed,
including schemes that employ Booth encoding~\cite{wahid20042D}
and the expansion factor method~\cite{TCSVT,JLPEA}.

\section{Improved AI based \mbox{1-D} DCT algorithm}

\begin{figure*}
\centering
\input{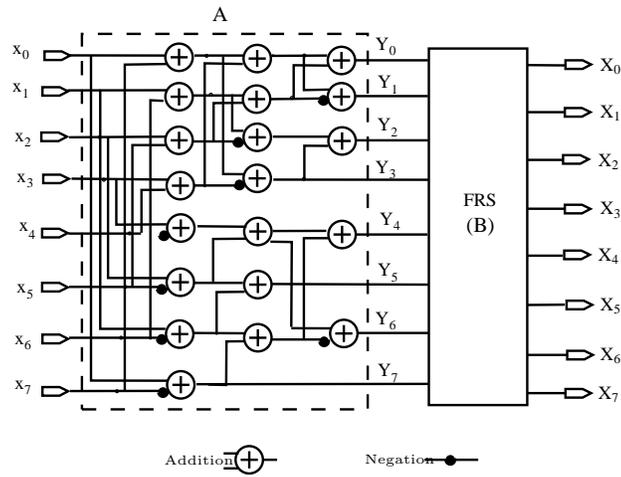}
\caption{Fast algorithm for AI based \mbox{1-D} DCT.}
\label{1D-DCT}
\end{figure*}

The proposed \mbox{1-D} DCT algorithm is derived from the
\mbox{1-D} AI Arai DCT~\cite{wahid20042D,TCSVT},
using algebraic manipulations.
Its computational complexity consists of 20 additions;
no multiplication or shift operations are required,
as depicted in Fig.~\ref{1D-DCT}.
This provides an improvement in terms of hardware resources
when compared with the algorithm proposed
in~\cite{TCSVT} and~\cite{wahid20042D},
which requires 21~additions and 2~shift operations.
Although when only a single DCT is considered,
the economy of one addition and two bit-shifting operations may
seem modest, we note that in video processing systems, the \mbox{1-D}
DCT is performed many times (once per column, per row, per component,
per 8$\times$8 block, per frame).
Thus,
the overall
computational savings is cumulative.
Further, the proposed method as well as its competitors are highly optimized
procedures; thus one may not expect enormous gains in terms of computational
complexity.
Indeed, we are approaching
the asymptotic limits of the theoretical DCT complexity.
Finally, it is noted that the savings in multiplications account for about
5\% in total complexity for considering DCTs only, which can be a significant gain
especially for low-power applications.

For the purpose of mathematical analysis,
the transform is described in matrix notation.
Let $\mathbf{A}$ denote the matrix transformation
associated to the
\mbox{1-D} Arai DCT algorithm defined over the AI structure
and
$\mathbf{B}$ denote the matrix related
to the operations in the FRS.
Matrix~$\mathbf{A}$ is represented
in terms of signal flow diagram
in the dashed block of Fig.~\ref{1D-DCT}.
Matrix~$\mathbf{B}$ is simply represents
the AI presentation,
ensuring that
the resulting calculation furnishes
AI as shown in~\eqref{z4.representation}.

Therefore,
the complete \mbox{1-D} AI DCT is given by
\begin{equation*}
\label{1d_eq1}
\mathbf{X}_{\text{1D}}
=
\mathbf{B}
\cdot
\mathbf{A}
\cdot
\mathbf{x}_{\text{1D}},
\end{equation*}
where $\mathbf{x}_{\text{1D}}$ and $\mathbf{X}_{\text{1D}}$
are 8-point column vectors for the input and output sequences, respectively,
and where
\begin{align}
\mathbf{A}
&=
\left[
\begin{smallmatrix}
   1 &   1 &   1 &   1 &   1 &   1 &   1 &   1 \\
   1 &  -1 &  -1 &   1 &   1 &  -1 &  -1 &   1 \\
   1 &   1 &  -1 &  -1 &  -1 &  -1 &   1 &   1 \\
   1 &   0 &   0 &  -1 &  -1 &   0 &   0 &   1 \\
   1 &   1 &   1 &   1 &  -1 &  -1 &  -1 &  -1 \\
   0 &  -1 &  -1 &   0 &   0 &   1 &   1 &   0 \\
  -1 &  -1 &   1 &   1 &  -1 &  -1 &   1 &   1 \\
   1 &   0 &   0 &   0 &   0 &   0 &   0 &  -1 \\
\end{smallmatrix}
\right]
,
\nonumber
\\
\label{FRS_matrix}
\mathbf{B}
&=
\left[
\begin{smallmatrix}
1 & 0 & 0 & 0 & 0 & 0 & 0 & 0 \\
0 & 1 & 0 & 0 & 0 & 0 & 0 & 0 \\
0 & 0 & 1 & z_1z_2 & 0 & 0 & 0 & 0 \\
0 & 0 & 1 & -z_1z_2 & 0 & 0 & 0 & 0 \\
0 & 0 & 0 & 0 & -z_2 & -z_1z_2 & -z_1 & 1 \\
0 & 0 & 0 & 0 & z_2 & -z_1z_2 & z_1 & 1 \\
0 & 0 & 0 & 0 & -z_1 & z_1z_2 & z_2 & 1 \\
0 & 0 & 0 & 0 & z_1 & z_1z_2 & -z_2 & 1
\end{smallmatrix}
\right]
.
\end{align}

The multiplications present in~$\mathbf{B}$
can be efficiently implemented
by means of the expansion factor method
as described in~\cite{JLPEA}.
The expansion factor method employs a factor $\alpha$
which scales the required multiplicands
$z_1$, $z_2$, and $z_1z_2$
into quantities that are close to integers governed by
the following relationship:
\begin{equation}
\label{eq_exp}
\alpha \cdot \begin{bmatrix} z_1 & z_2 & z_1z_2 \end{bmatrix}
\approx \begin{bmatrix}m_1 & m_2 & m_3\end{bmatrix}
,
\end{equation}
where $m_1$, $m_2$, and $m_3$ are integers.
This approach entail a reduction in the overall
number of multiplications required by the FRS~\cite{JLPEA}.

\section{Row-parallel \mbox{2-D} DCT architecture}

Let
$\mathbf{x}_\text{2D}$
and
$\mathbf{X}_\text{2D}$
be
8$\times$8 input and output
data of the \mbox{2-D} DCT,
respectively.
The \mbox{2-D} DCT of $\mathbf{X}_\text{2D}$
can be
obtained after
(i)~applying the \mbox{1-D} DCT to its columns;
(ii)~transposing the resulting matrix;
and
(iii)~applying the \mbox{1-D} DCT to the rows of the transposed matrix.
Mathematically,
above procedure is given by:
\begin{align}
\label{equation-X-2D}
\mathbf{X}_\text{2D}
=
\mathbf{B}\cdot\mathbf{A}
\cdot
\mathbf{x}_\text{2D}
\cdot
(\mathbf{B}\cdot\mathbf{A})^\top
=
\mathbf{B}\cdot\mathbf{A}
\cdot
\mathbf{x}_\text{2D}
\cdot
\mathbf{A}^\top
\!
\cdot
\mathbf{B}^\top
.
\end{align}

Matrix multiplications by $\mathbf{B}$ (reconstruction step)
should be the final computation stage.
Otherwise,
an earlier multiplication by $\mathbf{B}$
represents an intermediate reconstruction stage,
which may reintroduce numerical representation errors.
Such errors would then be propagate to subsequent
\mbox{1-D} DCT calls.
This may render the purpose of AI encoding ineffective.

 In~\cite{TCSVT} an AI based architecture,
where the reconstruction step occurs
only at the end of the computation,
was proposed.
In other words,
the reconstruction step was placed
after both row- and column-wise transforms.

\begin{figure*}
\centering
\scalebox{0.85}{
\input{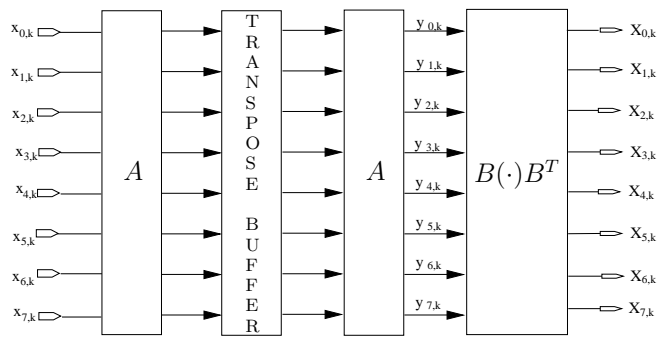}
}
\caption{Block diagram of the proposed \mbox{2-D} DCT architecture.}
\label{2D-DCT}
\end{figure*}

In the current contribution,
we propose a similar scheme.
However,
unlike
the architecture described in~\cite{TCSVT},
which requires a column-wise DCT call
for each of the four AI components,
we
propose an
architecture which
requires a single column-wise DCT.
As a result,
the computational complexity is greatly decreased in
this particular section of the algorithm.
The complete block diagram of the \mbox{2-D} DCT
architecture is given in Fig.~\ref{2D-DCT}.

\subsection{Derivation of the 1-Channel Architecture}

In this subsection,
we describe mathematically the derivation of the new
architecture.
Consider the following decomposition of the matrix $\mathbf{B}$,
which follows from~\eqref{FRS_matrix}:
\begin{align}
\label{equation-decomposition}
\mathbf{B}
=
\mathbf{B}_0 +
\mathbf{B}_1 \cdot z_1 +
\mathbf{B}_2 \cdot z_2 +
\mathbf{B}_3
\cdot z_1z_2,
\end{align}
where
\begin{align*}
\mathbf{B}_0
=
&
\left[
\begin{smallmatrix}
1 & 0 & 0 & 0 & 0 & 0 & 0 & 0 \\
0 & 1 & 0 & 0 & 0 & 0 & 0 & 0 \\
0 & 0 & 1 & 0 & 0 & 0 & 0 & 0 \\
0 & 0 & 1 & 0 & 0 & 0 & 0 & 0 \\
0 & 0 & 0 & 0 & 0 & 0 & 0 & 1 \\
0 & 0 & 0 & 0 & 0 & 0 & 0 & 1 \\
0 & 0 & 0 & 0 & 0 & 0 & 0 & 1 \\
0 & 0 & 0 & 0 & 0 & 0 & 0 & 1
\end{smallmatrix}
\right]
,
\quad
\mathbf{B}_1
=
\left[
\begin{smallmatrix}
0 & 0 & 0 & 0 & 0 & 0 & 0 & 0 \\
0 & 0 & 0 & 0 & 0 & 0 & 0 & 0 \\
0 & 0 & 0 & 0 & 0 & 0 & 0 & 0 \\
0 & 0 & 0 & 0 & 0 & 0 & 0 & 0 \\
0 & 0 & 0 & 0 & 0 & 0 & -1 & 0 \\
0 & 0 & 0 & 0 & 0 & 0 & 1 & 0 \\
0 & 0 & 0 & 0 & -1 & 0 & 0 & 0 \\
0 & 0 & 0 & 0 & 1 & 0 & 0 & 0
\end{smallmatrix}
\right]
\\
\mathbf{B}_2
=
&
\left[
\begin{smallmatrix}
0 & 0 & 0 & 0 & 0 & 0 & 0 & 0 \\
0 & 0 & 0 & 0 & 0 & 0 & 0 & 0 \\
0 & 0 & 0 & 0 & 0 & 0 & 0 & 0 \\
0 & 0 & 0 & 0 & 0 & 0 & 0 & 0 \\
0 & 0 & 0 & 0 & -1 & 0 & 0 & 0 \\
0 & 0 & 0 & 0 & 1 & 0 & 0 & 0 \\
0 & 0 & 0 & 0 & 0 & 0 & 1 & 0 \\
0 & 0 & 0 & 0 & 0 & 0 & -1 & 0
\end{smallmatrix}
\right]
,
\mathbf{B}_3
=
\left[
\begin{smallmatrix}
0 & 0 & 0 & 0 & 0 & 0 & 0 & 0 \\
0 & 0 & 0 & 0 & 0 & 0 & 0 & 0 \\
0 & 0 & 0 & 1 & 0 & 0 & 0 & 0 \\
0 & 0 & 0 & -1 & 0 & 0 & 0 & 0 \\
0 & 0 & 0 & 0 & 0 & -1 & 0 & 0 \\
0 & 0 & 0 & 0 & 0 & -1 & 0 & 0 \\
0 & 0 & 0 & 0 & 0 & 1 & 0 & 0 \\
0 & 0 & 0 & 0 & 0 & 1 & 0 & 0
\end{smallmatrix}
\right]
.
\end{align*}
Matrices
$\mathbf{B}_0$, $\mathbf{B}_1$, $\mathbf{B}_2$, and $\mathbf{B}_3$
are sparse, and
contain only $0$, $1$, and $-1$, leading to low complexity.

Therefore,
applying~\eqref{equation-decomposition} into~\eqref{equation-X-2D}
the \mbox{2-D} DCT computation can be rewritten as follows.
Let~$\mathbf{Y}_\text{2D} = \mathbf{A} \cdot \mathbf{x}_\text{2D} \cdot \mathbf{A}^\top$.
Notice that the calculation of $\mathbf{Y}_\text{2D}$
requires no multiplications.
Then, we have:
\begin{align}
\label{2D_full}
\mathbf{X}_\text{2D}
=&
\mathbf{B}
\cdot
\mathbf{Y}_\text{2D}
\cdot
\mathbf{B}^\top
\\ \nonumber
=&
\sum_{i=0}^3
\sum_{j=0}^3
\mathbf{B}_i
\cdot
\mathbf{Y}_\text{2D}
\cdot
\mathbf{B}_j^\top
\cdot
z_1^m
z_2^n
\end{align}
where $m,n\in \{0,1,2\}$.
Therefore,
the evaluation of $\mathbf{X}_\text{2D}$
is highly dependent on the computation of the
structure
$\mathbf{B}_i \cdot ( \cdot ) \cdot \mathbf{B}_j^\top$.

\begin{figure*}
\centering
\input{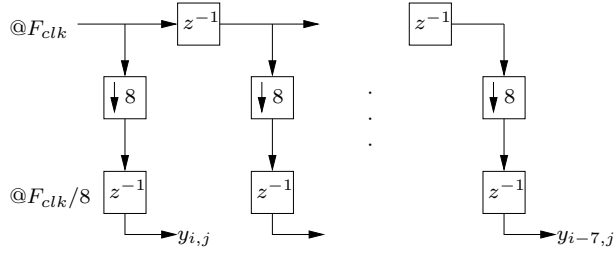}
\caption{SFG depicting the direct implementation of the buffer section in $\mathbf{B}_i \cdot ( \cdot ) \cdot \mathbf{B}_j^\top$ block for the $j$th row.}
\label{TB_SFG_8tap}
\end{figure*}

\section{Efficient Implementation of
$\mathbf{B}_i \cdot ( \cdot ) \cdot \mathbf{B}_j^\top$ block}

Matrices $\mathbf{B}_i$, $i\in \{ 0,1,2,3 \}$,
contain only a single non-zero element in each row.
Therefore any matrix multiplication by $\mathbf{B}_i$
can be trivially performed.

To maintain the row-parallel structure of the design,
the block that computes
$\mathbf{B}_i \cdot \mathbf{Y}_\text{2D}\cdot \mathbf{B}_j^\top$
should output a single row of data per clock cycle.
Hence the 64~elements in the 8$\times$8 $\mathbf{Y}_\text{2D}$ matrix should be stored for 8~clock cycles.
The signal flow graph~(SFG) corresponding to the $j$th row
of the storage structure is depicted in Fig~\ref{TB_SFG_8tap}.

The
left- and right-multiplication
of $\mathbf{Y}_\text{2D}$
by an arbitrary 8$\times$8 matrix
would require 56 storage elements operating at a clock rate of $F_\text{clk}$ and another 64 storage elements operating at a rate of $F_\text{clk}/8$.
However,
this operation requires
far less storage elements
due to the sparse nature of the matrices under consideration.
In the next section we investigate this possibility.

\subsection{Derivation of Half Column Independence}

The 8$\times$8 matrices
$\mathbf{B}_i$, $i\in \{ 0,1,2,3 \}$,
can be decomposed into a block matrix
of 4$\times$4 blocks in the following manner:
\begin{equation}
\mathbf{B}_i =
\begin{bmatrix}
\mathbf{B}_{i,0} & \mathbf{0}_4 \\
\mathbf{0}_4 & \mathbf{B}_{i,1}
\end{bmatrix}
,
\end{equation}
where $\mathbf{0}_4$ is the 4$\times$4 null matrix.
$\mathbf{B}_i$ is a block diagonal matrix.
Similarly,
\begin{align*}
\mathbf{Y}_\text{2D}
=
\begin{bmatrix}
\mathbf{Y}_{\text{2D},0} & \mathbf{Y}_{\text{2D},1} \\
\mathbf{Y}_{\text{2D},2} & \mathbf{Y}_{\text{2D},3}
\end{bmatrix}
.
\end{align*}
Therefore,
the product
$\mathbf{B}_i \cdot \mathbf{Y}_\text{2D} \cdot \mathbf{B}_j^\top$
is given by
\begin{align}\label{half_column}
\mathbf{B}_i \cdot \mathbf{Y}_\text{2D} \cdot \mathbf{B}_j^\top
=&
\begin{bmatrix}
\mathbf{B}_{i,0} & \mathbf{0}_4 \nonumber \\
\mathbf{0}_4 & \mathbf{B}_{i,1}
\end{bmatrix}
\cdot
\begin{bmatrix}
\mathbf{Y}_{\text{2D},0} & \mathbf{Y}_{\text{2D},1} \nonumber \\
\mathbf{Y}_{\text{2D},2} & \mathbf{Y}_{\text{2D},3}
\end{bmatrix}
\cdot
\\
&
\begin{bmatrix}
\mathbf{B}_{j,0}^\top & \mathbf{0}_4 \nonumber \\
\mathbf{0}_4 & \mathbf{B}_{j,1}^\top
\end{bmatrix}
\\
=&
\begin{bmatrix}
\mathbf{B}_{i,0} \cdot \mathbf{Y}_{\text{2D},0} \cdot \mathbf{B}_{j,0}^\top &
\mathbf{B}_{i,0} \cdot \mathbf{Y}_{\text{2D},1} \cdot \mathbf{B}_{j,1}^\top \\
\mathbf{B}_{i,1} \cdot \mathbf{Y}_{\text{2D},2} \cdot \mathbf{B}_{j,0}^\top &
\mathbf{B}_{i,1} \cdot \mathbf{Y}_{\text{2D},3} \cdot \mathbf{B}_{j,1}^\top
\end{bmatrix}
.
\end{align}

\begin{figure*}
\centering
\input{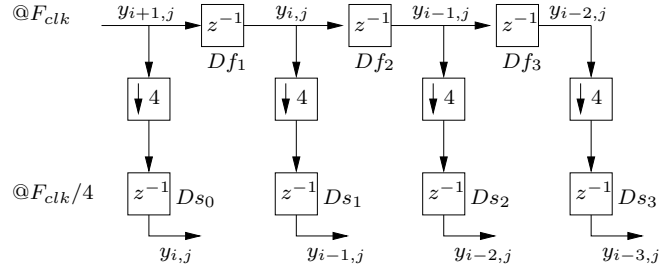}
\caption{SFG depicting the multiplexed implementation of the buffer section in $\mathbf{B}_i \cdot ( \cdot ) \cdot \mathbf{B}_j^\top$ block for the $j^{th}$ row.}
\label{TB_SFG_4tap}
\end{figure*}

Above relations show
that the computation of
$\mathbf{B}_i \cdot \mathbf{Y}_\text{2D} \cdot \mathbf{B}_j^\top$,
for a particular choice of $i$ and $j$,
can be performed in two sequential steps which can be \emph{computed independently} of each other. The computation of (\ref{half_column}) is achieved by the following steps:

\begin{enumerate}

\item
compute the first four rows of the resulting matrix using the first four rows of $\mathbf{Y}_\text{2D}$;

\item
compute the last four rows of the resulting matrix
using the last four rows of $\mathbf{Y}_\text{2D}$.

\end{enumerate}
The above sequence of computation can be realized using the SFG in
Fig.~\ref{TB_SFG_4tap}.
Eight consecutive rows are stored in clock FIFO registers. A total of 24 FIFO registers are needed for the section of the circuit operating at $F_\text{clk}$, while 32 registers are required for the slower section operating at $F_\text{clk}/8$.

\begin{figure*}
\centering
\scalebox{0.9}{
\input{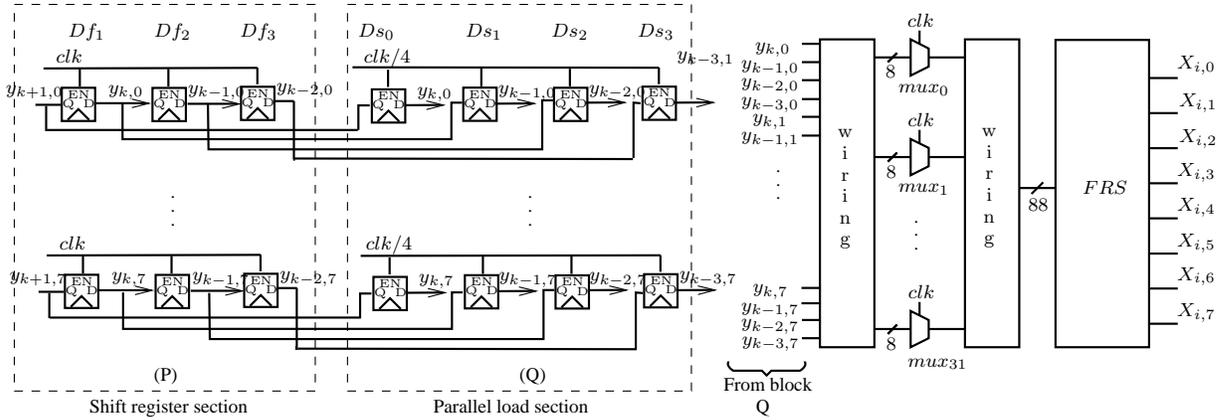}}
\caption{Implementation of $\mathbf{B(\cdot )B^T}$ block (Register content after the rising edge of $i$th clock cycle, where $i\pmod{4} = 0$ is shown).}
\label{TB}
\end{figure*}

\begin{figure*}
\centering
\scalebox{0.6}{\input{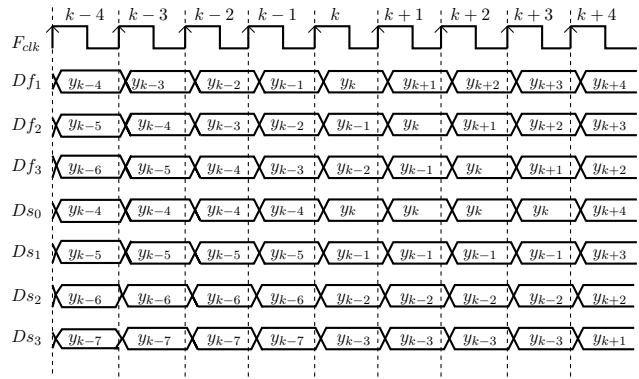}}
\caption{Timing diagram depicting the register content at each clock cycle of $F_\text{clk}$ with $k\pmod{4} = 0$.
Delay elements $Df_{\alpha},Ds_{\alpha}$, where $\alpha = 0,1,2,3$ are defined in Fig~\ref{TB_SFG_4tap}~(Active high, edge triggered).}
\label{4tap_timing_diag}
\end{figure*}

\subsection{Buffer}

The buffer shown in Fig.~\ref{TB} consists of a shift register section and a bank of parallel-load registers. At time
instant $k$,
the shift registers hold four consecutive rows of $\mathbf{Y}_\text{2D}$ denoted by
$y_{k,0\dots7}$,
$y_{k-1,0\dots7}$,
$y_{k-2,0\dots7}$,
and
$y_{k-3,0\dots7}$.
When $k\pmod{4} = 0$, a parallel load is executed, synchronously transferring the content of the shift registers~(Fig.~\ref{TB}, block P) into the parallel load register bank~(Fig.~\ref{TB}, block Q).
The timing diagram portraying register transfers in this block is shown in Fig~\ref{4tap_timing_diag}.
Transfers occur at every positive clock edge of $F_\text{clk}$.
Next, the computation of $\mathbf{B}_{i,q} \cdot \mathbf{Y}_{\text{2D},r} \cdot \mathbf{B}_{j,s}^\top$ block,
where
$i\in\{0,1,2,3\}$,
$q\in\{0,1\}$,
$r\in\{0,1,2,3\}$,
and
$s\in\{0,1\}$ is considered.

\subsection{Computing
$\mathbf{B}_{i,j} \cdot \mathbf{Y}_{\mathrm{2D},r} \cdot \mathbf{B}_{j,k}^\top$ Block}

The computation of
$\mathbf{B}_{i,j} \cdot \mathbf{Y}_{\text{2D},r} \cdot \mathbf{B}_{j,k}^\top$
is achieved by cross-wiring.
This step of the computation is free of addition and subtraction operations.
Eight-input multiplexers are used for cross-wiring.
Notice that
(i)~$\mathbf{B}_{i,j}\mathbf{Y}_{2D,0}$
and
$\mathbf{B}_{i,j}\mathbf{Y}_{2D,2}$
and
(ii)~$\mathbf{B}_{i,j}\mathbf{Y}_{2D,1}$
and
$\mathbf{B}_{i,j}\mathbf{Y}_{2D,3}$
are to be computed in a time synchronous scheme.
Thus, we need in total 32
multiplexers in total in order to compute all the 16 terms of (\ref{2D_full}).
The multiplexers are commuted time-synchronously,
leading to periodic connections being made to
the appropriate outputs of Block~Q in Fig.~\ref{TB}.
The wiring of the multiplexer inputs is governed by~(\ref{2D_full}) and (\ref{half_column}).

\subsection{Final Reconstruction Step (FRS)}

The cross-wiring form the $\mathbf{B}_i \cdot \mathbf{Y}_{\text{2D}} \cdot \mathbf{B}_j^\top$ block
are presented to
the FRS stage in order to recover the fixed point coefficients.
The computation is completely error free up to the FRS stage.
The FRS implementation based on expansion factors $\alpha$ = 4.5958 and 167.2309
proposed in~\cite{TCSVT}
was employed in the proposed hardware implementation.
These expansion factors correspond to integer sets $\begin{bmatrix}m_1 & m_2 & m_3\end{bmatrix} = \begin{bmatrix}12&5&13\end{bmatrix}$ and $\begin{bmatrix}437&181&473\end{bmatrix}$ respectively, with each set satisfying (\ref{eq_exp}).
The architecture using expansion factor 167.2309 offers a significant improvement in accuracy when compared to expansion factor 4.5958~\cite{TCSVT}.
In~\cite[Sec.~IV]{TCSVT}
a comprehensive account of the FRS implementation is furnished.
Such previously described FRS is applied in the proposed architecture.
\begin{table*}
\centering
\caption{Success rates of the DCT coefficient computation for various fixed-point bus widths and tolerance levels}
\label{success_rate}
\begin{tabular}{|c|c|c|c|c|c|c|c|c|}
\cline{2-9}
\multicolumn{1}{c|}{} & \multicolumn{8}{c|}{Percentage Tolerence}\tabularnewline
\hline
($m_1,m_2,m_3$) & $L$ & 10\% & 5\% & 1\% & 0.1\% & 0.05\% & 0.01\% & 0.005\%\tabularnewline
\hline
$(12,5,13)$ & 4 & 99.0367 & 98.1356 & 90.4067 & 51.7311 & 42.5511 & 31.4689 & 24.5189\tabularnewline
\cline{2-9}
 & 8 & 99.0356 & 98.0089 & 90.4433 & 51.9189 & 42.7911 & 31.4267 & 24.3556\tabularnewline
\hline
$(437,181,473)$ & 4 & 99.9867 & 99.9744 & 99.8856 & 98.9044 & 97.9 & 89.8322 & 80.8699\tabularnewline
\cline{2-9}
 & 8 & 99.99 & 99.9744 & 99.8856 & 98.9044 & 97.9 & 89.8322 & 80.8689\tabularnewline
\hline
\end{tabular}

\end{table*}

\section{Hardware Implementation and Results}

Two designs corresponding to expansion factors
4.5958 and 167.2309 were physically implemented and tested on-chip using field programmable gate array~(FPGA) technology and thereafter mapped to 45 nm CMOS technology. We employed a Xilinx ML605 evaluation kit which is populated with a
a Xilinx Virtex-6 XC6VLX240T FPGA device.
The JTAG interface was used to input the test 8$\times$8 \mbox{2-D} DCT arrays to the device from the \textsc{Matlab} workspace.
The measured outputs were returned to the \textsc{Matlab} workspace.

\subsection{On-chip Verification using Success Rates}

As a figure of merit,
we considered the success rate
defined as the percentage of coefficients
which are within the error limit of $\pm e\,\%$.
For the range $e = \{ 0.005, 0.01, 0.05, 0.1, 1, 5, 10 \}$,
the success rates
were of precision as given in the Table~\ref{success_rate}.
Input word length~$L$ was set to 4 and 8~bits.
The proposed AI architectures for the FPGA, are designed to be overflow-free at each stage throughout the AI encoded structure.
The accuracy results obtained are exactly same as the ones for the designs proposed in~\cite{TCSVT} based on expansion factor FRS.
Notice however that
up to the FRS all AI encoded 2-D DCT computation
is totally error free.
\begin{table*}
\centering
\caption{Area-speed and power consumption for FPGA implementation on Xilinx Virtex-6 XC6VLX240T}
\label{results_FPGA}
\begin{tabular}{|c|c|c|c|c|c|c|c|c|c|c|c|}
\hline
($m_1,m_2,m_3$) & $L$ &
\multicolumn{2}{c|}{Slices} &
\multicolumn{2}{c|}{Frequency~(MHz)} &
\multicolumn{2}{c|}{Dynamic power~(mW)} &
\multicolumn{2}{c|}{$AT$~($\text{slices}\cdot\mu \mathrm{s}$)} &
\multicolumn{2}{c|}{$AT^2$~($\text{slices} \cdot \mu \mathrm{s}^2$)}\tabularnewline
\cline{3-12}
 &  &~\cite{TCSVT} & Proposed &~\cite{TCSVT} & Proposed &~\cite{TCSVT} & Proposed &~\cite{TCSVT} & Proposed &~\cite{TCSVT} & Proposed\tabularnewline
\hline
$(12,5,13)$ & 4 & 2377 & 2212 & 309.9 & 316.8 & 1871 & 616 & 7.67 & 6.98 & 0.025 & 0.022\tabularnewline
\cline{3-12}
 & 8 & 3144 & 2536 & 300.4 & 294.3 & 1687 & 695 & 10.45 & 8.61 & 0.034 & 0.029\tabularnewline
\hline
$(437,181,473)$ & 4 & 2605 & 2291 & 312.4 & 312.9 & 912& 592 & 8.34& 7.32 & 0.028 & 0.024\tabularnewline
\cline{3-12}
 & 8 & 3445 & 2591 & 307.8 & 303.5 & 1123 & 714 & 11.19& 8.54 & 0.036 & 0.028\tabularnewline
\hline
\end{tabular}

\end{table*}

\begin{table*}
\centering
\caption{Simulated results for area-speed and power consumption from CMOS 45nm ASIC place and route~($V_\text{DD} = 1.1~\mathrm{V}$)}
\label{results_ASIC}
\begin{tabular}{|c|c|c|c|c|c|c|c|c|c|}
\hline
($m_1,m_2,m_3$)  & $L$ & Area  & Gate & Speed & \multicolumn{3}{c|}{Power(mW)} & $AT$ & $AT^2$\tabularnewline
\cline{6-8}
 &  & ($\mathrm{mm}^2$) & count & (MHz) & Leakage & Dynamic & Total & ($\mathrm{mm}^2\cdot \mathrm{ns}$)  & ($\mathrm{mm}^2\cdot \mathrm{ns}^2$) \tabularnewline
\hline
$(12,5,13)$  & 4 & 0.303 & 161.1K & 951 & 1.693 & 941.8 & 943.5 & 0.317 & 0.335\tabularnewline
\cline{3-10}
 & 8 & 0.404 & 215.2K & 949 & 2.220 & 1273.9 & 1276 & 0.426 & 0.449\tabularnewline
\hline
$(437,181,473)$  & 4 & 0.394 & 209.9K & 947 & 2.222 & 1073.7 & 1076 & 0.416 & 0.439\tabularnewline
\cline{3-10}
 & 8 & 0.439 & 233.9K & 946 & 2.942 & 1451.8 & 1455 & 0.464 & 0.490\tabularnewline
\hline
\end{tabular}

\end{table*}

\subsection{Area, Critical Path Delay, and Power Metrics}

The design was simulated up to place and route in 45~nm CMOS process using NCSU 45~nm PDK for ASICs and physically implemented using 40~nm CMOS Xilinx Virtex-6 XC6VLX240T FPGA.
Then the area~($A$),
critical path delay~($T$) and power consumption for each implementation were obtained, and are shown in Tables~\ref{results_FPGA} and~\ref{results_ASIC} for FPGA-based physical implementation and ASIC simulation.
The ASIC was simulated for power consumption and timing at an operating voltage~($V_\text{DD}$) of 1.1~V. Area utilization in $\mathrm{mm}^2$ and the gate count in terms of 2-input NAND gates are also provided.
The metrics were measured for different choices of finite precision using input word length $L \in\{4,8\}$ bits.
Similar metrics reported in~\cite{TCSVT} using the same FPGA device and tools are also given in Table~\ref{results_FPGA} for comparison purposes.

The area utilization in FPGA implementation is given in Table~\ref{results_FPGA} in terms of the number of elementary programmable logic blocks~(slices) whereas the ASIC equivalent is given in terms of the chip area.
The total power consumption of the hardware design is constituted of static~(leakage) and dynamic components.
Static power consumption in FPGAs is dominated by leakage power of the logic fabric and of the configuration memory. Thus this quantity is essentially independent of the proposed design.
Hence, only the dynamic power consumption of the FPGA implementation is given. For the ASIC implementation both leakage and dynamic power components are given in Table~\ref{results_ASIC}.
The area-time~($AT$) and area-time-squared~($AT^2$) metrics are also provided as a measurement of the overall performance.
The $AT$ performance is a suitable metric for area efficient designs while the $AT^2$ metric could be used for designs with the speed of operation as the optimization goal~\cite{thompson1979area}.
The proposed architecture leads to a reduction of 23\% of $AT$ and 22\% of
$AT^2$~(Table~\ref{results_FPGA}) when compared with the architecture requiring five 1-D DCT cores~\cite{TCSVT}, for designs using the set of integers
$\begin{bmatrix}437&181&473\end{bmatrix}$
with 8-bit input word length.
The CMOS design with clock frequency of 951~MHz amounts to
a pixel rate of 7.608~G pixels/sec and
an 8$\times$8 block rate of 118.875~M blocks/sec.

\subsection{Comparison with Other Architectures}

Table~\ref{comparison} provides a comparison between the existing AI based \mbox{2-D} DCT architectures and designs based on the  proposed architecture. Eight-bit versions of designs using expansion factors
4.5958 and 167.2309 corresponding to integer sets
$\begin{bmatrix}12&5&13\end{bmatrix}$
and
$\begin{bmatrix}437&181&473\end{bmatrix}$,
respectively,
are used for the comparison.
The comparison clearly indicates the advantages of the proposed scheme with respect to throughput accuracy and flexibility.

\begin{table*}
\centering
\footnotesize
\caption{Comparison of the proposed implementation with existing algebraic
integer implementations}
\label{comparison}

\begin{tabular}{|c|c|c|c|c|c|c|c|}
\cline{2-8}
\multicolumn{1}{c|}{} & Nandi et al. & Jullien et al. & Wahid et al. & \multicolumn{2}{c|}{Madanayake et al.~\cite{TCSVT}} & \multicolumn{2}{c|}{Proposed architectures}\tabularnewline
\cline{5-8}
\multicolumn{1}{c|}{} &~\cite{nandi} &~\cite{fu_2004} &~\cite{wahid_murtuza} & $(12,5,13)$  & $(437,181,473)$ & $(12,5,13)$  & $(437,181,473)$\tabularnewline
\hline
Measured results & No & No & No & Yes & Yes & Yes & Yes\tabularnewline
\hline
 & Single \mbox{1-D} & Two \mbox{1-D} DCT & Two \mbox{1-D} DCT & Five \mbox{1-D} DCT & Five \mbox{1-D} DCT & Two \mbox{1-D} DCT & Two \mbox{1-D} DCT\tabularnewline
Structure & DCT +Mem. & +Dual port & +TMEM & +TMEM & +TMEM & +TMEM & +TMEM\tabularnewline
 & bank & RAM &  &  &  &  & \tabularnewline
\hline
Multipliers & 0 & 0 & 0 & 0 & 0 & 0 & 0\tabularnewline
\hline
Exact 2D AI  & No & No & No & Yes & Yes & Yes & Yes\tabularnewline
computation &  &  &  &  &  &  & \tabularnewline
\hline
Operating  &  &  &  &  &  &  &\tabularnewline
frequency & N/A & 75 & 194.7 & 300.39 & 307.78 & 949 & 946\tabularnewline
(MHz) &  &  &  &  &  &  & \tabularnewline
\hline
8$\times$ 8 blocks  &  &  &  &  &  &  &\tabularnewline
per clock cycle & 1/128 & 1/64 & 1/64 & 1/8 & 1/8 & 1/8 & 1/8\tabularnewline
\hline
8$\times$8 Block rate & 7.8125 & 1.171 & 3.042 & 37.55 & 38.47 & 118.625 & 118.25\tabularnewline
($\times 10^6s^{-1}$) &  &  &  &  &  &  & \tabularnewline
\hline
Pixel rate  & 125 & 75 & 194.7 & 1158.95 & 1187.35 & 7592 & 7568\tabularnewline
($\times 10^6s^{-1}$) &  &  &  &  &  &  & \tabularnewline
\hline
Implementation  & Xilinx  &
0.18~$\mathrm{\mu m}$ &
0.18~$\mathrm{\mu m}$ &
Xilinx &
Xilinx & 45~nm & 45~nm\tabularnewline
technology & XC5VLX30 & CMOS & CMOS & XCVLX240T & XCVLX240T & CMOS & CMOS\tabularnewline
\hline
Coupled  &  &  &  &  &  &  & \tabularnewline
quantization & Yes & Yes & Yes & No & No & No & No\tabularnewline
noise &  &  &  &  &  &  & \tabularnewline
\hline
Independantly  &  &  &  &  &  &  & \tabularnewline
adjustable & No & No & No & Yes & Yes & Yes & Yes\tabularnewline
precision &  &  &  &  &  &  & \tabularnewline
\hline
FRS between  &  &  &  &  &  &  & \tabularnewline
row-column & No & Yes & Yes & No & No & No & No\tabularnewline
stages &  &  &  &  &  &  & \tabularnewline
\hline
\end{tabular}
\end{table*}

\section{Conclusion}

An area efficient row-parallel architecture for 8$\times$8 \mbox{2-D} DCT computation based on AI number representation leading to exact computations up to the FRS is proposed.
The number of integer channels after the column-wise transform is reduced to one down from four in~\cite{TCSVT} by eliminating the redundancies present in AI channels.
This further enables the simplification of  the \mbox{1-D} blocks and the overall hardware complexity where only two \mbox{1-D} DCT blocks are needed.
Architectural variants corresponding to two expansion factors for the FRS were physically implemented, each at 4-bit and 8-bit input precision, and subsequently verified on a Xilinx Virtex-6 XC6VLX240T FPGA device.
The maximum clock frequency for the FPGA realizations is  294.3 MHz.  In addition, the architectures were  mapped to custom silicon realizations using 45~nm CMOS technology.
The realizations were simulated up to place and route stage but no fabrications were attempted.
The simulated designs achieved a
potential maximum clock frequency of 951~MHz as
reported by the Cadence Encounter tool.

\bibliographystyle{ieeetr}
\bibliography{references}

\end{document}